\documentclass [10pt,a4paper]{article}
\usepackage[numbers,sort&compress]{natbib}
\usepackage{amsfonts}
\usepackage{indentfirst}       
\usepackage{amsmath}           
\usepackage{amsfonts}          
\usepackage{amssymb}           
\usepackage{amsthm}            
\usepackage{maplestd2e}
\usepackage{a4wide}
\usepackage[CJKbookmarks]{hyperref}
\usepackage{graphicx}
\usepackage{float}
\usepackage{subfigure}

\def\Rmn#1{\expandafter\uppercase\expandafter{\romannumeral #1}}

\begin{document}
\newtheorem{theorem}{Theorem}[section]
\newtheorem{definition}{Definition}[section]
\newtheorem{lemma}{Lemma}[section]
\newtheorem{corollary}{Corollary}[section]
\title{Simplex  Subdivisions and  Nonnegativity Decision of
Forms\footnote{ Partially supported by a National Key Basic Research
Project of China (2004CB318000) and by National Natural Science
Foundation of China (10571095). Sponsored by K.C.Wong Magna Fund in
Ningbo University. }}
\date{}

\author{
Xiaorong Hou$^1$~~Song Xu$^{1,2}$  \\
 \textit{\small $1.$ College of Automation,
University of Electronic Science and Technology of China, Sichuan,
PRC}\\
 \textit{\small $2.$ Faculity of Science, Ningbo University, Ningbo,
Zhejiang, PRC}\\
\textit{\small E-mail:
\href{mailto:houxr@uestc.edu.cn}{houxr@uestc.edu.cn},
\href{mailto:xusong@nbu.edu.cn}{xusong@nbu.edu.cn} }}

\maketitle
{\noindent\small  {\bf Abstract}
 This paper mainly studies  nonnegativity decision
of forms based on variable substitutions.  Unlike  existing
research, the paper regards  simplex subdivisions  as  new
perspectives  to study variable substitutions,  gives some
subdivisions of the simplex $\mathbb{T}_n$, introduces the concept
of convergence of the subdivision
 sequence,  and presents a suf\mbox{}ficient and
necessary condition for the convergent self-similar subdivision
sequence. Then the relationships between subdivisions and their
corresponding  substitutions are established. Moreover,  it is
proven that if the form $F$ is indef\mbox{}inite on $\mathbb{T}_n$
and the sequence of the successive $L$-substitution sets is
convergent, then the sequence of sets
$\{\textrm{SLS}^{(m)}(F)\}_{m=1}^\infty$ is negatively terminating,
and  an algorithm for deciding indef\mbox{}inite forms with a
counter-example  is obtained. Thus, various effective
 substitutions  for deciding positive semi-definite forms
and indefinite forms  are gained, which are beyond the weighted
dif\mbox{}ference substitutions
 characterized by ``difference''.}\\[2ex]

\\[2ex]


\section{ Introduction }
  Theories and methods of nonnegative polynomials have been  widely
used in  robust control, non-linear control and non-convex
optimization \cite{P:1,P:2,J:1}, etc.   Some famous research works
on nonnegativity decision of polynomials without cell-decomposition
were given  by P\'{o}lya's Theorem \cite{Po:1,G:1} and papers
\cite{Ca:1,H:2}.

  Recently, Yang \cite{Yang:1,Yang:2,Yang:3} introduced a heuristic
method for nonnegativity  decision of polynomials, which is now
called Successive Dif\mbox{}ference Substitution (SDS).  It has been
applied to prove a great many polynomial inequalities with more
variables and higher degrees.  Yao  \cite{Yong1} investigated the
weighted dif\mbox{}ference
 substitutions instead of the  original dif\mbox{}ference
 substitutions, and proved that, for a
form (namely, a homogeneous polynomial) which is positive definite
on $\mathbb{R}_+^n$, the corresponding sequence of SDS sets is
positively terminating, where
$\mathbb{R}_+^n=\{(x_1,x_2,...,x_n)|x_i\geq 0, i=1,2,...,n\}.$ That
is,  we can decide the nonnegativity of a positive definite form by
 successively running SDS finite times.
The  research results above are all
 confined to the variable substitutions characterized by
``difference''.

 Unlike  existing research, this paper
regards  simplex subdivisions  as  new perspectives  to study
variable substitutions, and obtains various effective variable
substitutions  for deciding positive semi-definite forms and
indefinite forms on $\mathbb{R}_+^n$, which are beyond the weighted
dif\mbox{}ference substitutions characterized by ``difference''. The
paper is organized as follows. Section 2 gives some subdivisions of
the simplex $\mathbb{T}_n$, and establishes the relationships
between them and their corresponding  substitutions. Section 3
introduces the concept of  the termination of
$\{\textrm{SLS}^{(m)}(F)\}_{m=1}^\infty$, which is directly related
to the positive semi-definite property of the form $F$. Section 4
proves that if the form $F$ is indef\mbox{}inite on $\mathbb{R}_+^n$
and  the sequence of the successive $L$-substitution sets is
convergent, then the sequence of sets
$\{\textrm{SLS}^{(m)}(F)\}_{m=1}^\infty$ is negatively terminating.
An algorithm for deciding an indef\mbox{}inite form with a
counter-example is presented in Section 5, and by using the obtained
algorithm, several examples are listed in Section 6.

\section{Simplex
subdivisions and the corresponding substitutions }

We first introduce some def\mbox{}initions and notations.
\begin{definition}
   \emph{Let $V=[v_{ij}]$ be an $n\times n$ matrix. If $\sum\limits_{i=1}^nv_{ij}=1, j=1,2,\cdots
   n$, $V$ is called a normalized matrix.  And the corresponding linear transformation
   \begin{equation}\label{fff}
X^{\textmd{Tr}}=VT^{\textmd{Tr}},
\end{equation}}
\emph{is called a normalized substitution,  where $X, T\in
\mathbb{R}_{+}^{n}$,   $X^{\textmd{Tr}}$,  $T^{\textmd{Tr}}$ are
respectively the transposes of $X$, $T$, and  $V$ is called the
substitution matrix of (\ref{fff}).}
    \end{definition}

\begin{lemma}  \label{zc}
\emph{Let $ U = [u_{ij}] = V_1 V_2 \cdots V_k$. If
$V_1,V_2,\cdots,V_k$ are all normalized matrices, then $U$ is a
normalized matrix , that is,
 \begin{displaymath}
 \sum\limits_{i=1}^nu_{ij}=1(j=1,2,\cdots n).
 \end{displaymath}
 And let
 $(x_1,x_2,\cdots,x_n)^{\textmd{Tr}}=U(t_1,t_2,\cdots,t_n)^{\textmd{Tr}}$,
 then
 $\sum\limits_{i=1}^{n}x_i=1$ iff
 $\sum\limits_{i=1}^{n}t_i=1$.}
\end{lemma}
 The proof of Lemma  \ref{zc} is very straightforward and is omitted.
\begin{definition}
   \emph{ If the form
   $F(X)\geq0$  for all $X\in\mathbb{R}_+^n$,  then $F$ is called positive semi-def\mbox{}inite on
   $\mathbb{R}_+^n$, and the set of all positive
semi-definite forms is denoted by PSD; If
   $F(X)>0$  for all $X(\neq0)\in\mathbb{R}_+^n$, then
   $F$ is called positive def\mbox{}inite on
   $\mathbb{R}_+^n$(briefly, a PD); If there are $X$ and $Y\in \mathbb{R}_+^n$,
such that $F(X)>0$ and  $F(Y)<0$,  then
   $F$ is called indef\mbox{}inite on
   $\mathbb{R}_+^n$. }
\end{definition}

 The $(n-1)$-dimensional simplex  is defined as follows
$$\mathbb{T}_n = \{(x_1,x_2,\cdots,x_n)|\sum\limits_{i=1}^nx_i=1 , (x_1,x_2,\cdots,x_n)\in\mathbb{R}_{+}^{n}\}.$$
\begin{definition}
   \emph{ If the form
   $F(X)\geq0$  for all $X\in\mathbb{T}_n$,  then $F$ is called positive semi-definite on
   $\mathbb{T}_n$;
    If
   $F(X)>0$  for all $X\in\mathbb{T}_n$, then
   $F$ is called positive definite on
   $\mathbb{T}_n$; If there are $X$ and $Y\in \mathbb{T}_n$
such that $F(X)>0$ and  $F(Y)<0$,  then
   $F$ is called indefinite on
   $\mathbb{T}_n$. }
\end{definition}

    It is easy to get the following conclusion.
 \begin{lemma}  \label{abcd}
\emph{The form $F$ is positive semi-definite (positive definite,
indefinite) on $\mathbb{R}_+^n$
    iff $F$ is positive semi-definite (positive definite,
indefinite) on
   $\mathbb{T}_n$.}
\end{lemma}

According to Lemma \ref{abcd}, for brevity, we suppose the form $F$
is defined on  $\mathbb{T}_n$ in the following sections.

\subsection{The
barycentric subdivision and the weighted dif\mbox{}ference
substitutions} In this subsection, we focus on   the barycentric
subdivision and its corresponding substitutions.
 \begin{figure}[H]
\begin{center}
\includegraphics[width=0.25\textwidth]{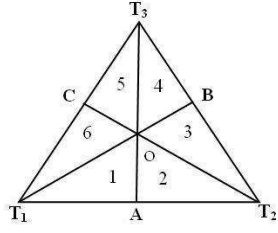}
 \caption{the barycentric subdivision}\label{Fig1}
\end{center}
\end{figure}
See Fig.1.  The first barycentric subdivision of the simplex
$\mathbb{T}_3$ consists of 6 subsimplexes.  Consider the subsimplex
$T_1AO$ labeled 1 first,  where $T_1=(1,0,0), A
=(\frac{1}{2},\frac{1}{2},0), O=(\frac{1}{3},
\frac{1}{3},\frac{1}{3})$ in  the $T_1T_2T_3$-coordinate system.
Consruct the following substitution
\begin{equation}\label{dh1}X^{\textmd{Tr}}=W_3T^{\textmd{Tr}},\end{equation}
where
\begin{equation}
W_3 = \left [ \begin{array}{c} T_1\\A\\O\end{array}\right
]^{\textmd{Tr}} =\left[
\begin{array}{lll}
 1 & \frac{1}{2} & \frac{1}{3} \\
0 & \frac{1}{2} & \frac{1}{3} \\
0 & 0 & \frac{1}{3}
\end{array} \right].
\end{equation}
By (\ref{dh1}), if $X=T_1=(1,0,0)$, then $T=(1,0,0)$; If
$X=A=(\frac{1}{2},\frac{1}{2},0)$, then $T=(0,1,0)$; If
$X=O=(\frac{1}{3},\frac{1}{3},\frac{1}{3})$ , then $T=(0,0,1)$.
  And it is indicated that  the subsimplex
$T_1AO$ and the substitution (\ref{dh1}) correspond to each other.

Analogously,  the other five subsimplexes labeled as 2-6  correspond
 to the five substitutions,  respectively,  whose substitution
  matrices are as follows
\begin{equation}
 \left[
\begin{array}{lll}
0 & \frac{1}{2} & \frac{1}{3} \\
1& \frac{1}{2} & \frac{1}{3} \\
0 & 0 & \frac{1}{3}
\end{array} \right],\left[
\begin{array}{lll}
0 & 0 & \frac{1}{3} \\
1 & \frac{1}{2} & \frac{1}{3} \\
0 & \frac{1}{2} & \frac{1}{3}
\end{array} \right],
\left[
\begin{array}{lll}
0 & 0 & \frac{1}{3} \\
0 & \frac{1}{2} & \frac{1}{3} \\
1 & \frac{1}{2} & \frac{1}{3}
\end{array} \right],
\left[
\begin{array}{lll}
0 & \frac{1}{2} & \frac{1}{3} \\
0 &  0 & \frac{1}{3} \\
1 & \frac{1}{2} & \frac{1}{3}
\end{array} \right],
\left[
\begin{array}{lll}
1 & \frac{1}{2} & \frac{1}{3} \\
0 & 0 & \frac{1}{3} \\
0 & \frac{1}{2} & \frac{1}{3}
\end{array} \right].
\end{equation}

The six substitutions above  are just  the weighted
dif\mbox{}ference substitutions
 for $n=3$, and the set which consists of all the
  substitutions is called the weighted
dif\mbox{}ference substitution set. Therefore,   the fisrt
barycentric subdivision of $ \mathbb{T}_3$ corresponds to the
weighted dif\mbox{}ference substitution set, that is, there is a
one-to-one correspondence between the subsimplexes of the first
barycentric subdivision of $ \mathbb{T}_3$ and the weighted
dif\mbox{}ference substitutions.
\begin{definition}\label{zcyydh}
\emph{ Let $PW_3$ be the weighted dif\mbox{}ference substitution
  matrix set for $n=3$.
 The
 set of
linear transformations
\begin{displaymath}
\{X^{\textmd{Tr}}=B_{[\alpha_1]}B_{[\alpha_2]}\cdots
B_{[\alpha_m]}T^{\textmd{Tr}}|B_{[\alpha_i]}\in PW_3\},
\end{displaymath}
is called the $m$-times successive weighted dif\mbox{}ference
substitution set, which consists of $6^m$  substitutions.}
\end{definition}

By Lemma \ref{zc} and Def\mbox{}inition \ref{zcyydh}, we have that
the $m$-th barycentric subdivision of the simplex $\mathbb{T}_3$
corresponds to the $m$-times successive weighted dif\mbox{}ference
substitution set.

\subsection{Some more  subdivisions  and the corresponding substitutions}

Next, we'll present some more  subdivisions of the simplex
$\mathbb{T}_3$  and their corresponding substitutions.
\begin{figure}[H]
  \centering
  \subfigure[]{
    \label{fig:subfig:a} 
    \includegraphics[width=0.235\textwidth]{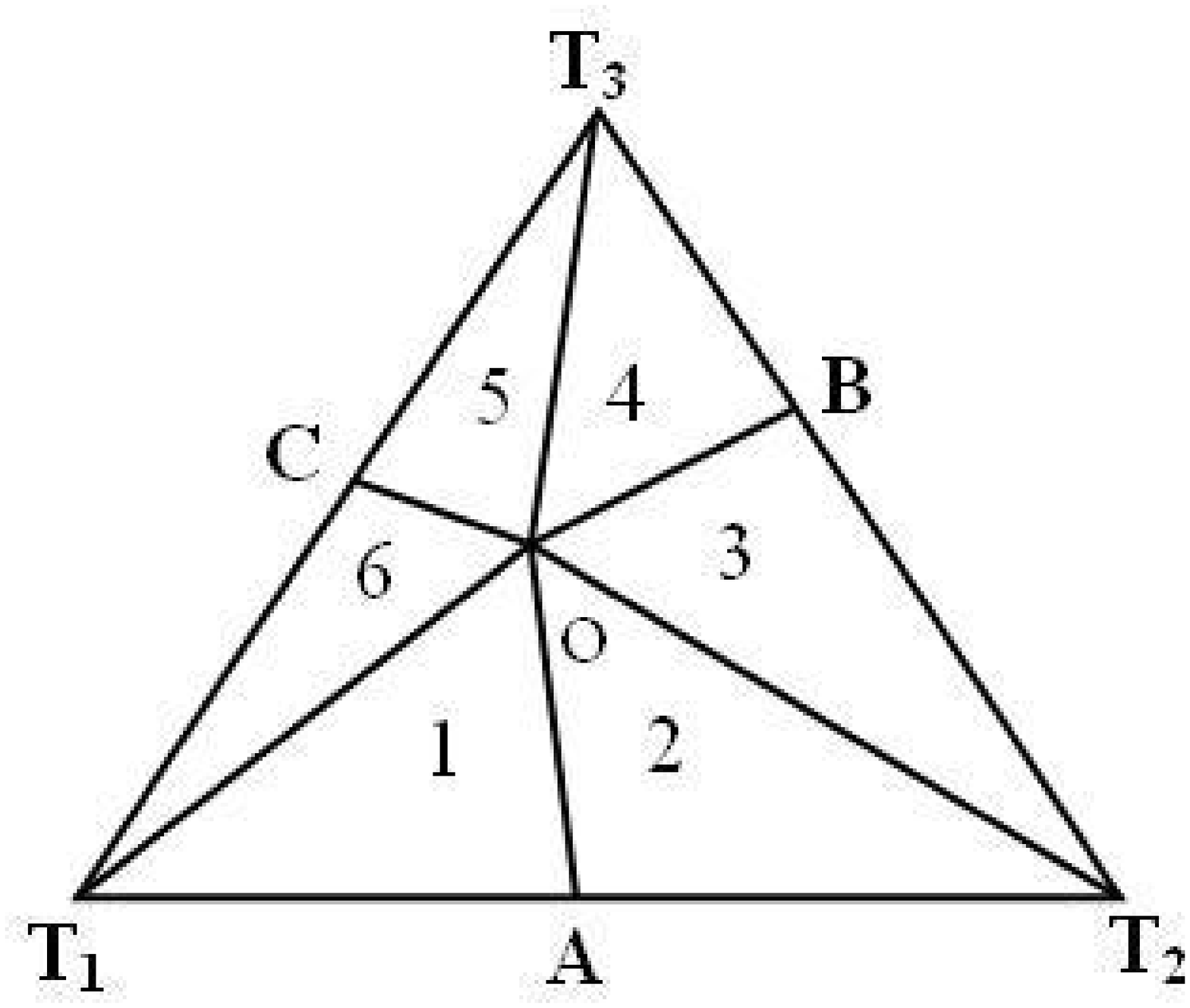}}
      \hspace{0.15in}
    \subfigure[]{
    \label{fig:subfig:b} 
    \includegraphics[width=0.20\textwidth]{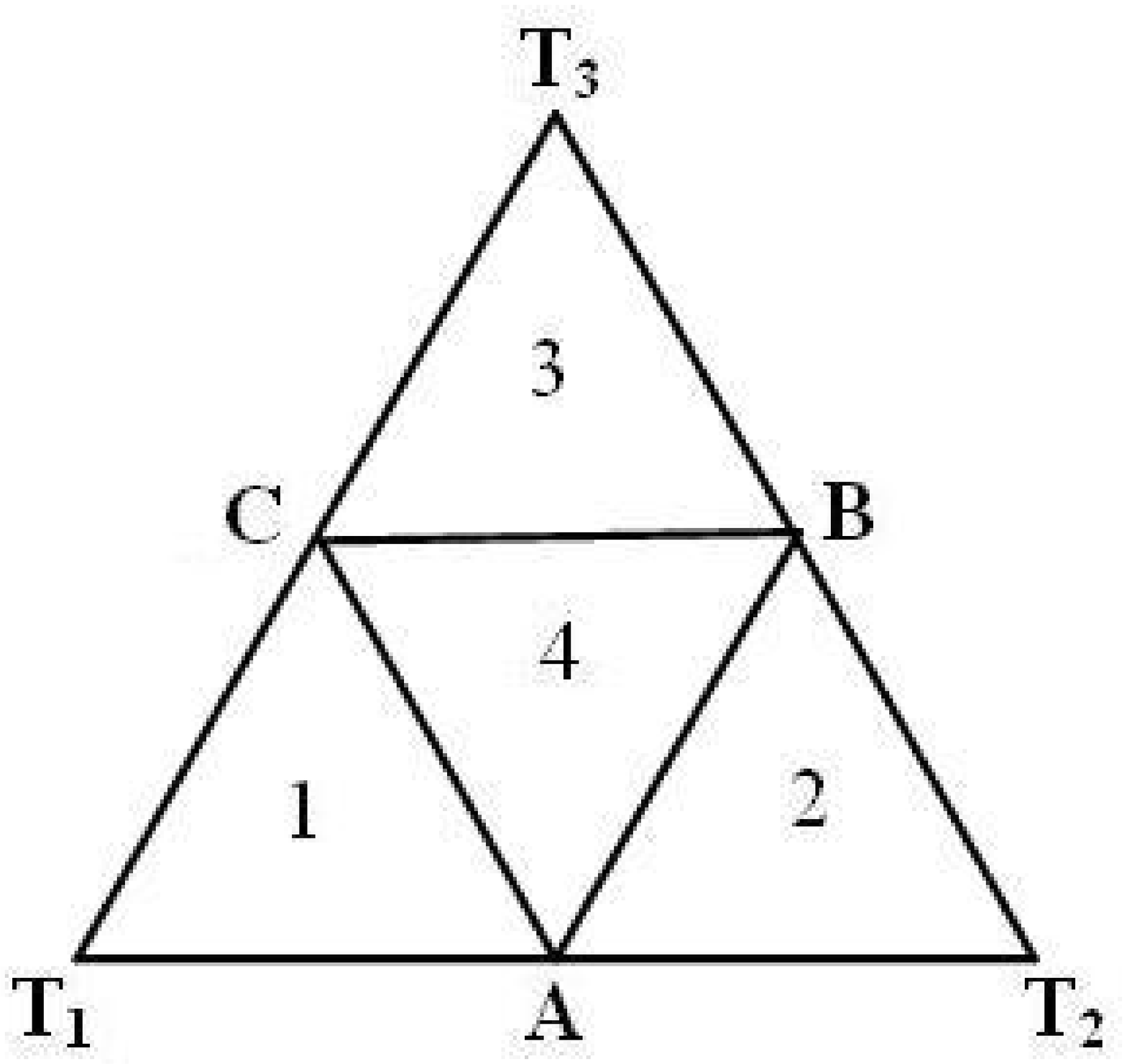}}
      \hspace{0.15in}
    \subfigure[]{
    \label{fig:subfig:c} 
    \includegraphics[width=0.20\textwidth]{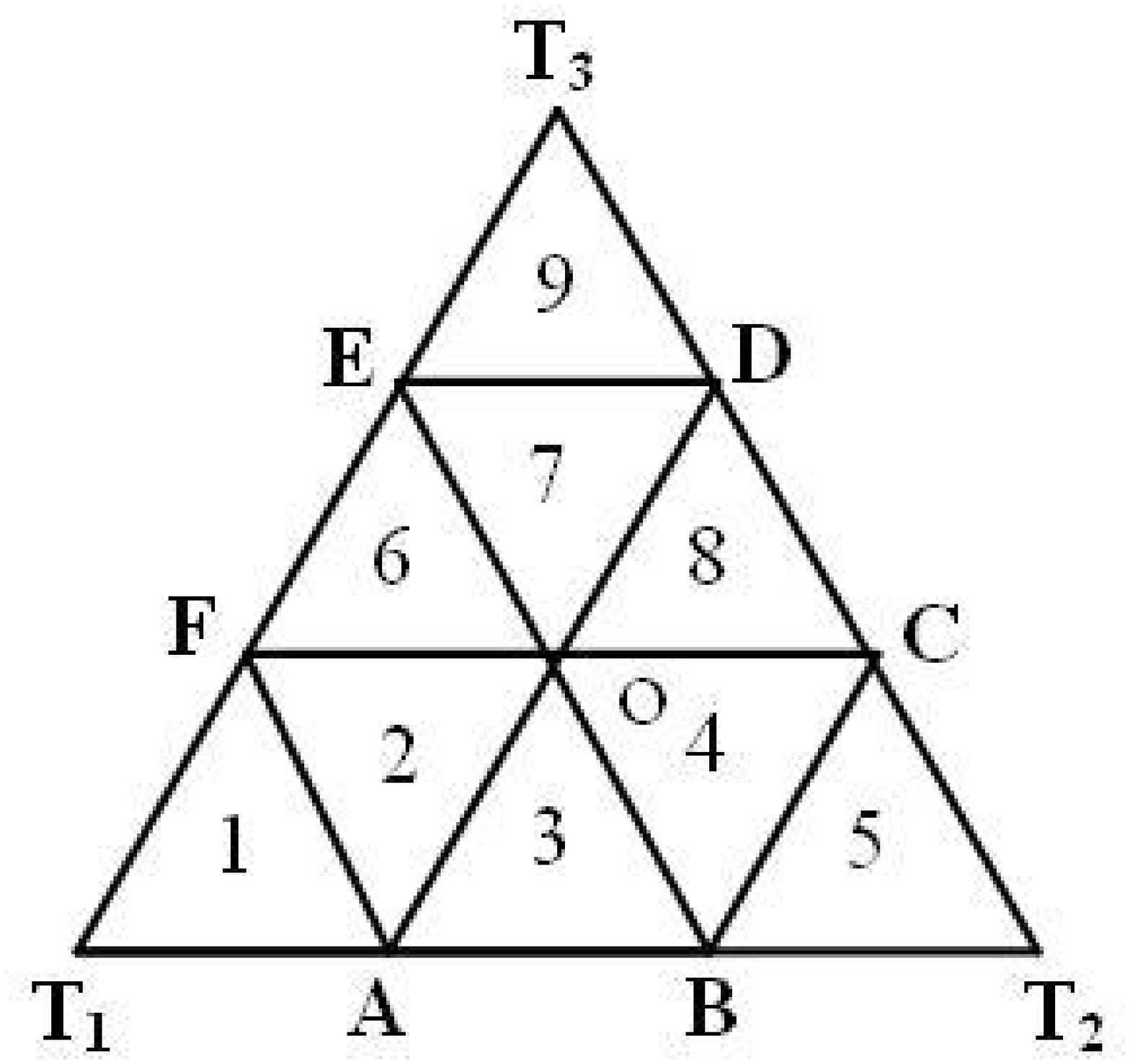}}
  \caption{Some   subdivisions}
  \label{fig:subfig} 
\end{figure}

 Consider  Fig.2(a) first. Let $T_1=(1,0,0)$,
$T_2=(0,1,0)$, $T_3=(0,0,1)$, $O=(a_0,b_0,c_0 )$, $A=(a_1 ,b_1,0 )$,
$B=(0 ,b_2,c_2 )$, $C=(a_3 ,0,c_3 )$. And the first subdivision of
the simplex $\mathbb{T}_3$ consists of six subsimplexes labeled as
1-6, which  correspond to the following substitution
  matrices, respectively.
\begin{equation}
\begin{split}
\left [ \begin{array}{c} T_1\\A\\O\end{array}\right ]^{\textmd{Tr}}
 =\left[
\begin{array}{lll}
1 & a_1 & a_0 \\
0 & b_1  & b_0 \\
0 & 0 & c_0
\end{array} \right],
\left [ \begin{array}{c} T_2\\A\\O\end{array}\right ]^{\textmd{Tr}}
= \left[
\begin{array}{lll}
0 & a_1 & a_0 \\
1 & b_1  & b_0 \\
0 & 0 & c_0
\end{array} \right],
\left [ \begin{array}{c} T_2\\B\\O\end{array}\right
]^{\textmd{Tr}}=\left[
\begin{array}{lll}
0 & 0 & a_0 \\
1 & b_2  & b_0 \\
0 & c_2 & c_0
\end{array} \right],\\
\left [ \begin{array}{c} T_3\\B\\O\end{array}\right
]^{\textmd{Tr}}=\left[
\begin{array}{lll}
0 & 0 & a_0 \\
0 & b_2  & b_0 \\
1 & c_2 & c_0
\end{array} \right],
\left [ \begin{array}{c} T_3\\C\\O\end{array}\right
]^{\textmd{Tr}}=\left[
\begin{array}{lll}
0 & a_3 & a_0 \\
0 & 0  & b_0 \\
1 & c_3 & c_0
\end{array} \right],
\left [ \begin{array}{c} T_1\\C\\O\end{array}\right
]^{\textmd{Tr}}=\left[
\begin{array}{lll}
1 & a_3 & a_0 \\
0 & 0  & b_0 \\
0 & c_3 & c_0
\end{array} \right].
\end{split}
\end{equation}
where $a_i\neq 0, b_i\neq 0$, and $ c_i\neq 0$. If $O=(\frac{1}{3},
\frac{1}{3}, \frac{1}{3})$, $A=(\frac{1}{2} ,\frac{1}{2} ,0 )$,
$B=(0 ,\frac{1}{2}, \frac{1}{2})$, $C=(\frac{1}{2} ,0,\frac{1}{2}
)$, then the subdivision is just  the barycentric subdivision.

In Fig.2(b), the first subdivision of the simplex $\mathbb{T}_3$
consists of four subsimplexes labeled as 1-4, which correspond to
the following substitution
  matrices, respectively.
\begin{equation}
\left[
\begin{array}{lll}
1 & \frac{1}{2} & \frac{1}{2} \\
0 & \frac{1}{2} & 0 \\
0 & 0 & \frac{1}{2}
\end{array} \right],
 \left[
\begin{array}{lll}
0 & 0 & \frac{1}{2} \\
1& \frac{1}{2} & \frac{1}{2} \\
0 & \frac{1}{2} & 0
\end{array} \right],\left[
\begin{array}{lll}
0 & \frac{1}{2} & 0 \\
0 & 0 & \frac{1}{2} \\
1 & \frac{1}{2} & \frac{1}{2}
\end{array} \right],
\left[
\begin{array}{lll}
\frac{1}{2} & 0 & \frac{1}{2} \\
\frac{1}{2}& \frac{1}{2} & 0 \\
0 & \frac{1}{2} & \frac{1}{2}
\end{array} \right].
\end{equation}

In Fig.2(c), the first subdivision of the simplex $\mathbb{T}_3$
consists of nine subsimplexes labeled as 1-9, which  correspond to
the following substitution
  matrices, respectively.
\begin{equation}
\begin{split}
\left[
\begin{array}{lll}
1 & \frac{2}{3} & \frac{2}{3} \\
0 & \frac{1}{3} & 0 \\
0 & 0 & \frac{1}{3}
\end{array} \right],
 \left[
\begin{array}{lll}
\frac{2}{3} & \frac{1}{3} & \frac{2}{3} \\
\frac{1}{3} & \frac{1}{3} & 0 \\
0 &  \frac{1}{3} & \frac{1}{3} \\
\end{array} \right],
\left[
\begin{array}{lll}
\frac{2}{3} & \frac{1}{3} & \frac{1}{3} \\
\frac{1}{3} & \frac{2}{3} & \frac{1}{3} \\
0 &  0 & \frac{1}{3} \\
\end{array} \right],
\left[
\begin{array}{lll}
\frac{1}{3} & 0 & \frac{1}{3} \\
\frac{2}{3} & \frac{2}{3} & \frac{1}{3} \\
0 & \frac{1}{3} & \frac{1}{3} \\
\end{array} \right],
\\\left[
\begin{array}{lll}
\frac{1}{3} & 0 & 0 \\
\frac{2}{3} & 1 & \frac{2}{3} \\
0 & 0 & \frac{1}{3}
\end{array} \right],
\left[
\begin{array}{lll}
\frac{2}{3} & \frac{1}{3} & \frac{1}{3} \\
0 & \frac{1}{3} & 0 \\
\frac{1}{3} & \frac{1}{3} & \frac{2}{3}
\end{array} \right],
 \left[
\begin{array}{lll}
\frac{1}{3} &0 & \frac{1}{3} \\
\frac{1}{3} & \frac{1}{3} & 0 \\
\frac{1}{3}&  \frac{2}{3} & \frac{2}{3} \\
\end{array} \right],
\left[
\begin{array}{lll}
\frac{1}{3} & 0 & 0 \\
\frac{1}{3} & \frac{2}{3} & \frac{1}{3} \\
\frac{1}{3} &  \frac{1}{3} & \frac{2}{3} \\
\end{array} \right],
\left[
\begin{array}{lll}
\frac{1}{3} & 0 & 0 \\
0 & \frac{1}{3} & 0 \\
\frac{2}{3} & \frac{2}{3} & 1 \\
\end{array} \right].
\end{split}
\end{equation}

Next, we'll give the definitions of the  self-similar subdivision
sequence and the successive  substitution set.

\begin{definition}\label{ddd}
\emph{Given a simplex $K$. Let $\mbox{sd}^0(K)=K$, and
$\mbox{sd}^i(K)=\mbox{sd}(\mbox{sd}^{i-1}(K))$ , where
$\mbox{sd}^i(K)$ is the subdivision of $\mbox{sd}^{i-1}(K)$ for
$i=1,2,\cdots$.  For an arbitrary subsimplex $\sigma$ of
$\mbox{sd}^i(K)$($i=0,1,\cdots$),  if the vertex coordinates of all
the subsimplexes of  $\mbox{sd}(\sigma)$ in the $\sigma$-coordinate
system  equal to the corresponding ones of all the subsimplexes of
 $\mbox{sd}(K)$ in the $K$-coordinate system, then
$\mbox{sd}^i(K)$ is called the $i$-th self-similar subdivision of
the simplex $K$, and the subdivision sequence
$\{\mbox{sd}^i(\mathbb{T}_n)\}_{i=1}^\infty$ is called
self-similar.}
\end{definition}

Fig.3 shows the first and second self-similar subdivisions of the
simplex $\mathbb{T}_3$.
\begin{figure}[H]
  \centering
  \subfigure[]{
    \label{fig:subfig:a2} 
    \includegraphics[width=0.20\textwidth]{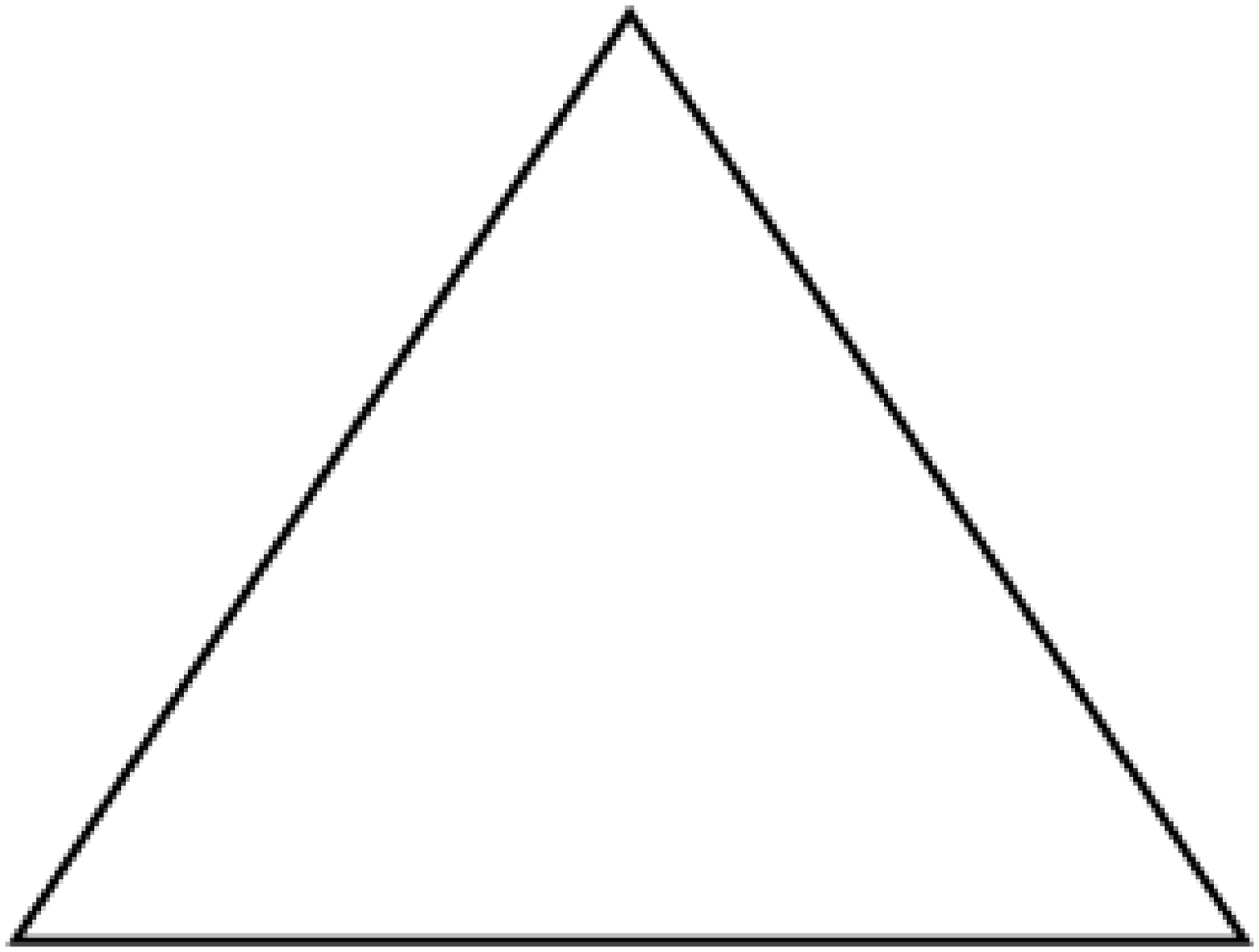}}
    \hspace{0.15in}
    \subfigure[]{
    \label{fig:subfig:b2} 
    \includegraphics[width=0.20\textwidth]{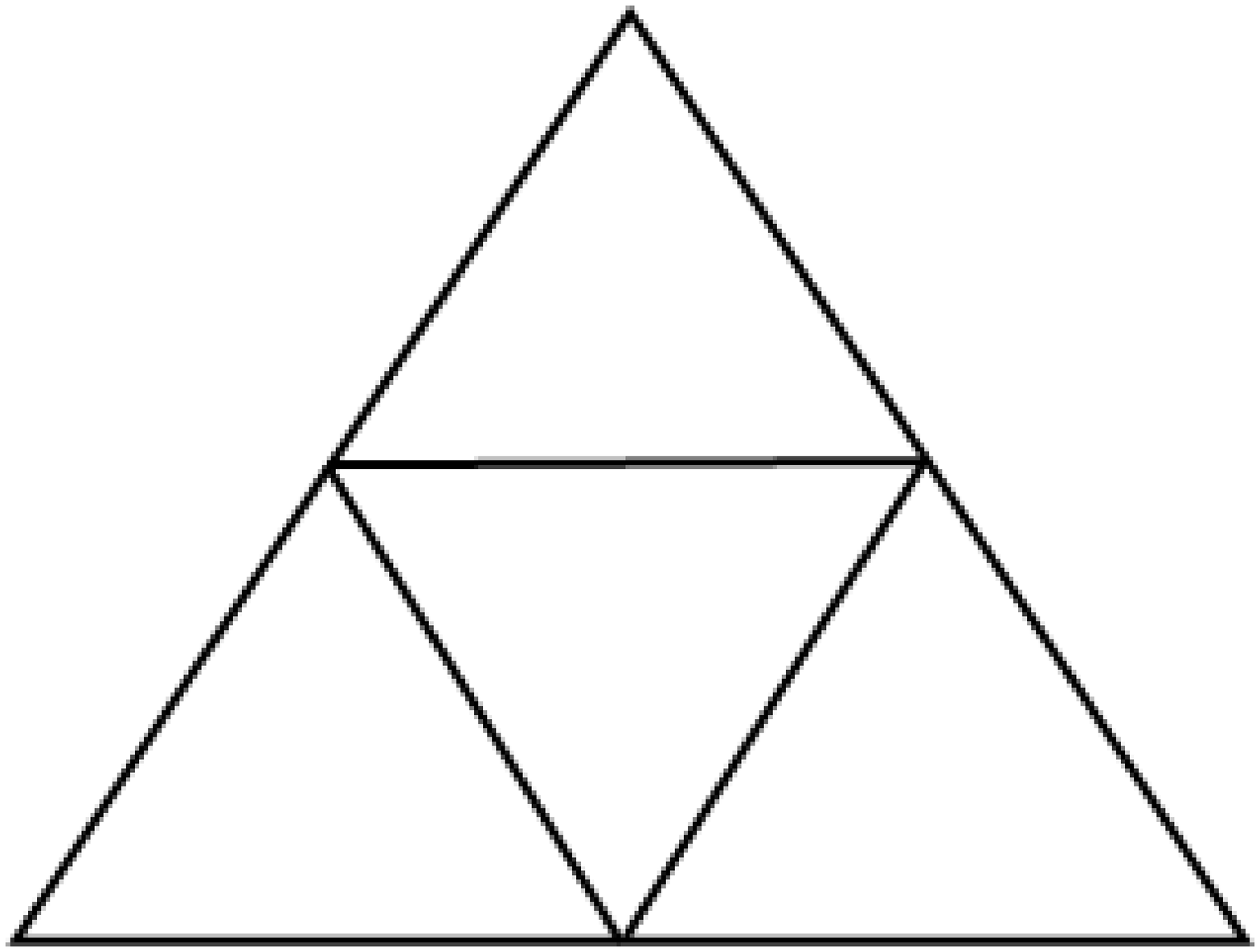}}
    \hspace{0.15in}
    \subfigure[]{
    \label{fig:subfig:c2} 
    \includegraphics[width=0.20\textwidth]{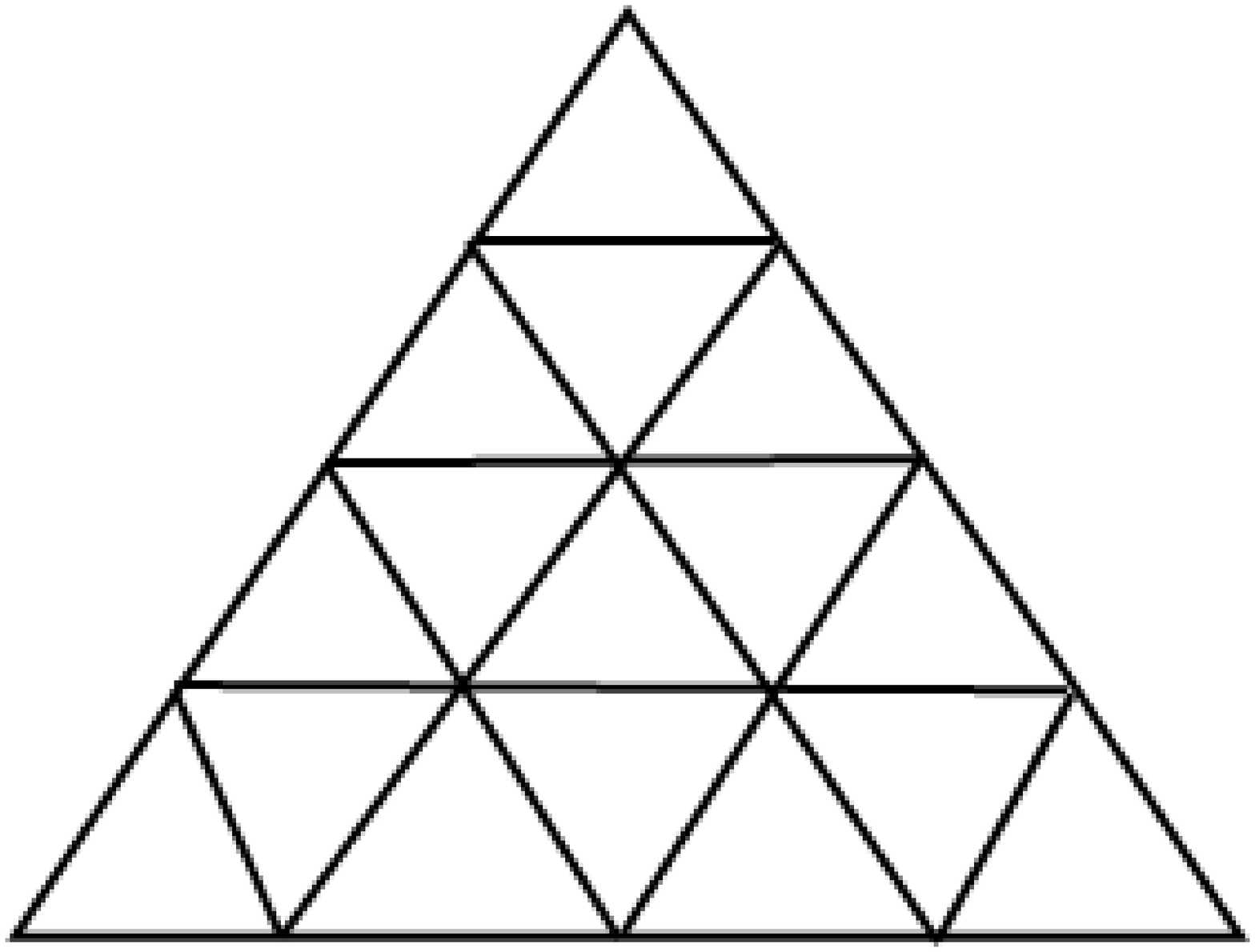}}
  \caption{the  self-similar subdivisions of
the simplex $\mathbb{T}_3$}
  \label{fig:subfig} 
\end{figure}

\begin{definition}\label{eee}
\emph{ Let $PA$ be the  $L$-substitution
  matrix set.
 The
 set of
linear transformations
\begin{displaymath}
\{X^{\textmd{Tr}}=A_{[1]}A_{[2]}\cdots
A_{[m]}T^{\textmd{Tr}}|A_{[i]}\in PA, i=1,2,\cdots,m\},
\end{displaymath}
is called the $m$-times successive   $L$-substitution  set, briefly,
the $m$-times \textmd{SLS} set.}
\end{definition}
Suppose that the the first subdivision of the simplex $\mathbb{T}_3$
corresponds to the normalized  $L$-substitution  set. By Lamma
\ref{zc}, Definition \ref{ddd} and Definition \ref{eee}, we conclude
 that the $m$-th self-similar subdivision of the simplex
$\mathbb{T}_3$ corresponds to the $m$-times successive
$L$-substitution
 set.

Easily,  all the subdivisions of the simplex   $\mathbb{T}_3$
 in the paper  can be expanded to the case of the $(n-1)$-dimensional simplex
 $\mathbb{T}_n$.

\subsection{Convergence of the subdivision
 sequence of a simplex }

In this subsection, we'll
 introduces the concept
of the  convergence of the subdivision
 sequence,  and presents a suf\mbox{}ficient and
necessary condition for the convergent self-similar subdivision
sequence.

\begin{definition}\label{sld1} \emph{ Let $\sigma$ be a subsimplex of $\mathbb{T}_n$,  the maximum distance between
 vertexs of $\sigma$
is called the diameter of $\sigma$.}
\end{definition}

\begin{definition}\label{sld0}
\emph{ Let $K$ and  $\mbox{sd}^i(K)$ be defined as  Definition
\ref{ddd}.  If for all $ \varepsilon>0$, there exists $N\in
\mathbb{N}$, such that all the diameters of the subsimplexes
 of $\mbox{sd}^N(K)$ are less than $\varepsilon$, the subdivision sequence
$\{\mbox{sd}^i(K)\}_{i=1}^\infty$ is called convergent.}
\end{definition}

\begin{definition}\label{sld5}
\emph{Suppose that the subdivision scheme through which
$\mbox{sd}^{i-1}(K)$ is subdivided into $\mbox{sd}^i(K)$ corresponds
to the substitution set $L_{i}$ for
 $i=1,2,\cdots$. If the subdivision sequence
$\{\mbox{sd}^i(K)\}_{i=1}^\infty$ is convergent, then the
 sequence of substitution sets
$\{L_i\}_{i=1}^\infty$ is called convergent, and if $L_{i}=L$,
$i=1,2,\cdots$, brief\mbox{}ly, we say that the sequence of the
successive $L$-substitution sets is convergent.}
\end{definition}

Next, we'll consider  the convergence of the sequence of the
successive weighted dif\mbox{}ference substitution sets.

\begin{lemma}\label{bs}
\emph{\cite{Edwin1,James1} Let $K$ be a complex. If $K_N$ is the
$k$-th barycentric subdivision of $K$, then for all $\varepsilon>0$,
there exists $N\in \mathbb{N}$, such that all the diameters of the
subsimplexes
 of $K_N$ are less than $\varepsilon$. }
\end{lemma}

 By Lemma \ref{bs} , we have the following theorem.
\begin{theorem}
\emph{The   barycentric subdivision
 sequence $\{K_i\}_{i=1}^\infty$ of $ \mathbb{T}_n$ and the corresponding sequence of
the successive weighted dif\mbox{}ference substitution sets
 are convergent.} \label{sl}
\end{theorem}
Now, we present a suf\mbox{}ficient and necessary condition for the
convergent self-similar subdivision sequence, which plays important
 roles for
   nonnegativity decision
of forms.

\begin{theorem}\label{zxs}
\emph{Let $\{\mbox{sd}^i(\mathbb{T}_n)\}_{i=1}^\infty$ be the
self-similar subdivision sequence of $\mathbb{T}_n$, then it is
convergent iff  an  arbitrary $1$-dimensional proper face of
subsimplexes of $\mbox{sd}(\mathbb{T}_n)$ is not the $1$-dimensional
one of $\mathbb{T}_n$. }
\end{theorem}

\begin{proof} The proof of necessity is straightforward by Definition \ref{ddd} and Definition \ref{sld0}.
Next,  we prove sufficiency. Let $d$ be the diameter of
$\mathbb{T}_n$, and $r$ be the maximum diameter of  subsimplexes of
$\mbox{sd}(\mathbb{T}_n)$, then $r\leq d$.  Since  an arbitrary
$1$-dimensional proper face of subsimplexes of
$\mbox{sd}(\mathbb{T}_n)$ is not the $1$-dimensional one of
$\mathbb{T}_n$, we have $r<d$. Note $l=\frac{r}{d}$, then $l<1$.
Obviously,  the maximum diameter of subsimplexes  of the $N$-th
self-similar subdivision of $\mathbb{T}_n$ is $l^Nd$. When
$N\longrightarrow\infty$, then $l^Nd\longrightarrow0$, so the
self-similar subdivision sequence
$\{\mbox{sd}^i(\mathbb{T}_n)\}_{i=1}^\infty$ is convergent.
\end{proof}
Consider the following subdivisions of $\mathbb{T}_3$ in Fig.4.

 In Fig.4(a), the first subdivision of the simplex
$\mathbb{T}_3$ consists of 3 subsimplexes  labeled as 1-3, which
correspond to the following substitution
  matrices, respectively.
\begin{equation}
\left[
\begin{array}{lll}
1 & 0 & \frac{1}{3} \\
0 & 1 & \frac{1}{3} \\
0 & 0 & \frac{1}{3}
\end{array} \right],
 \left[
\begin{array}{lll}
0 & 0 & \frac{1}{3} \\
1& 0 & \frac{1}{3} \\
0 & 1 & \frac{1}{3}
\end{array} \right], \left[
\begin{array}{lll}
0 & 1 & \frac{1}{3} \\
0& 0 & \frac{1}{3} \\
1 & 0 & \frac{1}{3}
\end{array} \right].
\end{equation}

In Fig.4(b), the first subdivision of the simplex $\mathbb{T}_3$
consists of 5 subsimplexes labeled as 1-5, which correspond to the
following substitution
  matrices, respectively.
\begin{equation}
\begin{split}
\left[
\begin{array}{lll}
1 & 0 & a_0 \\
0 & 1  & b_0 \\
0 & 0 & c_0
\end{array} \right],
 \left[
\begin{array}{lll}
0 & 0 & a_0 \\
1 & b_1  & b_0 \\
0 &c_1 & c_0
\end{array} \right],
 \left[
\begin{array}{lll}
0 & 0 & a_0 \\
0 & b_1  & b_0 \\
1 &c_1 & c_0
\end{array} \right],
\left[
\begin{array}{lll}
0 & a_2 & a_0 \\
0 & 0  & b_0 \\
1 & c_2 & c_0
\end{array} \right],
\left[
\begin{array}{lll}
1 & a_2 & a_0 \\
0 & 0  & b_0 \\
0 & c_2 & c_0
\end{array} \right] .
\end{split}
\end{equation}
where $O=(a_0,b_0,c_0 )$, $A=(a_1 ,b_1,0 )$, and  $B=(0 ,b_2,c_2 )$.

 In Fig.4, it is straightforward that the
$1$-dimensional proper face $T_1T_2$ of the subsimplex $T_1T_2O$ is
just the  one of $T_1T_2T_3$. By Theorem \ref{zxs},  we have that
the two self-similar subdivision sequences of $\mathbb{T}_n$ aren't
convergent.
\begin{figure}[H]
  \centering
  \subfigure[]{
    \label{fig:subfig:a} 
    \includegraphics[width=0.245\textwidth]{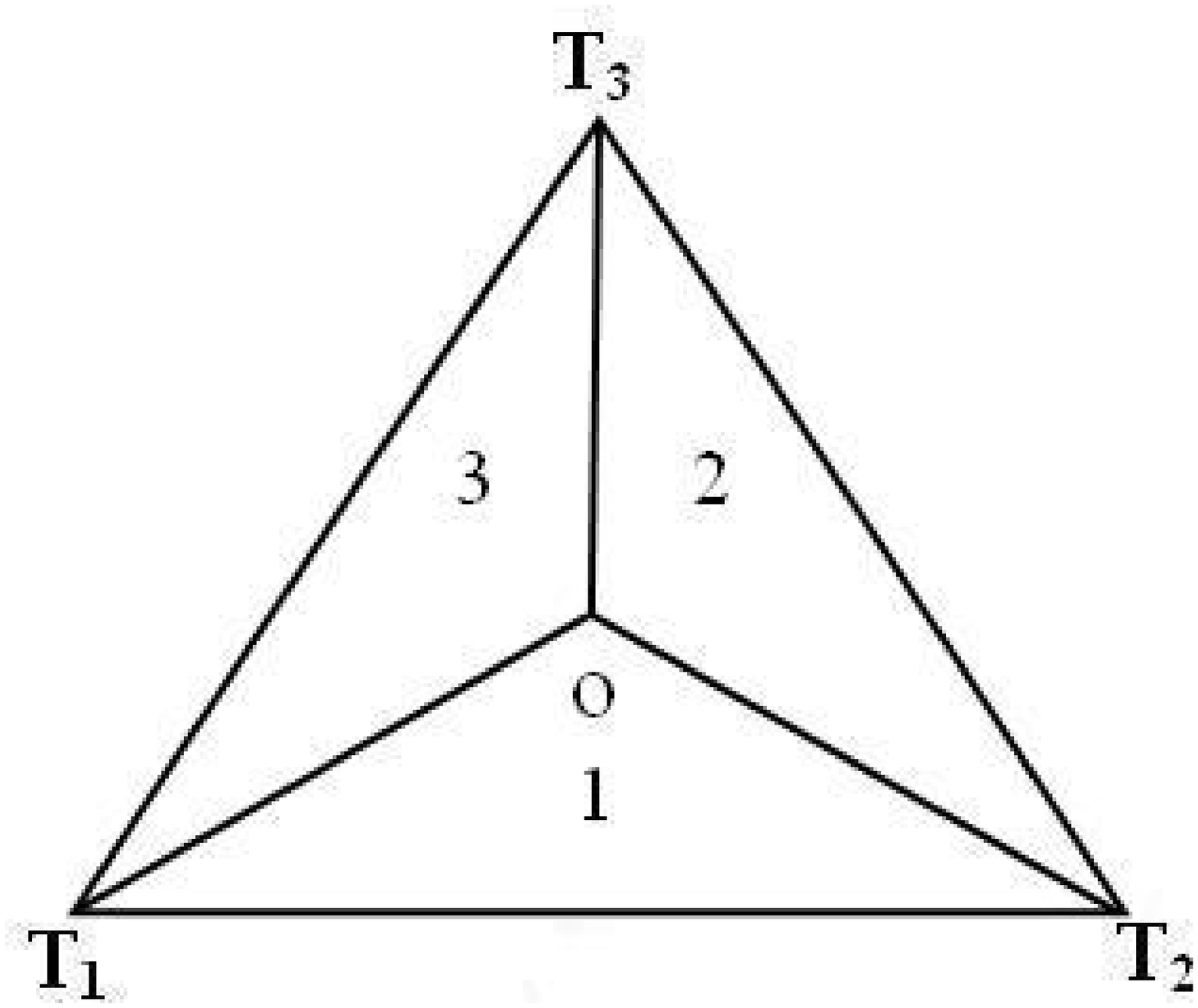}}
    \subfigure[]{
    \label{fig:subfig:b} 
    \hspace{0.15in}
    \includegraphics[width=0.245\textwidth]{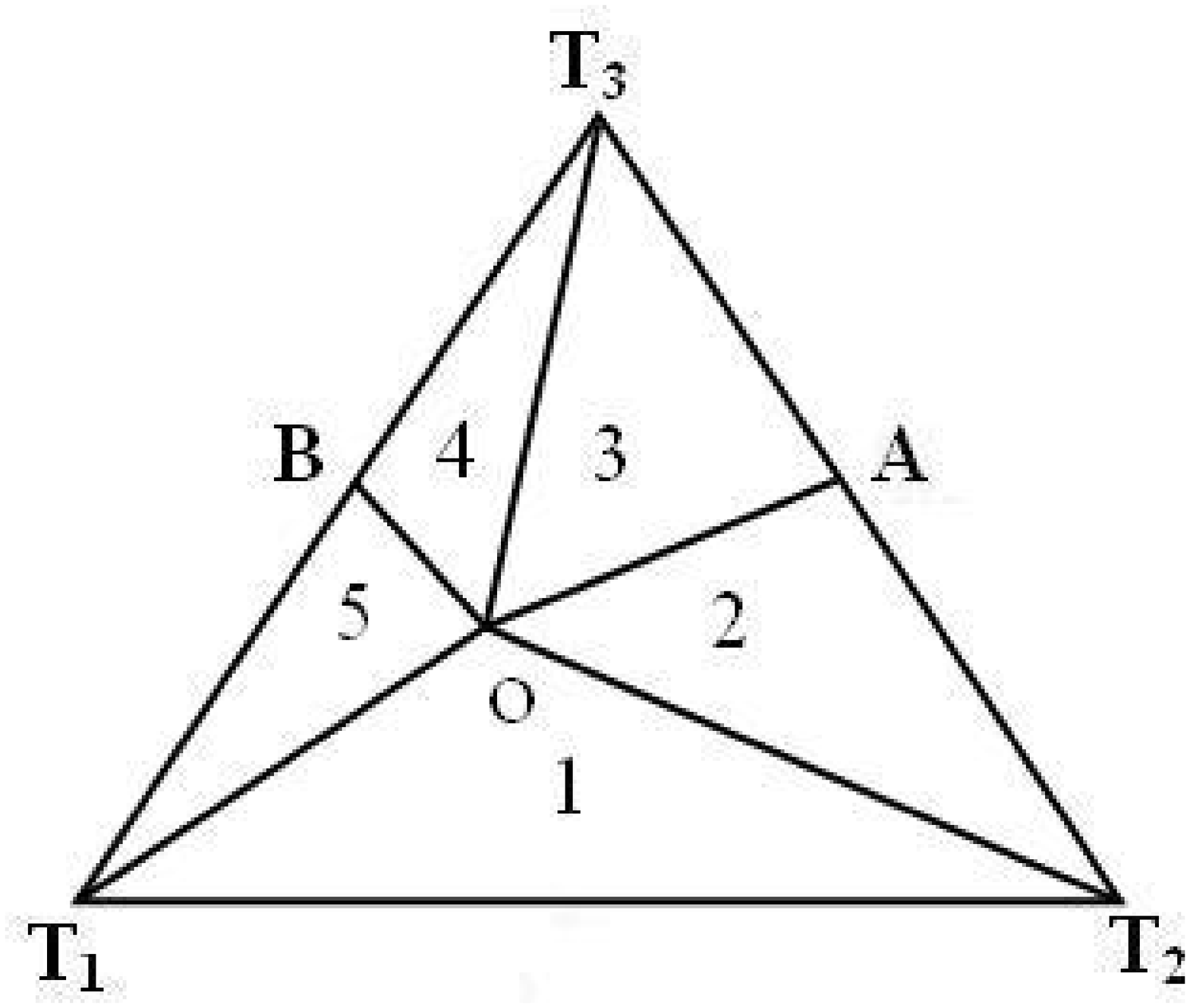}}
  \caption{Some more  subdivisions}
  \label{fig:subfig} 
\end{figure}

\section{Termination of the Sequence of the Successive
Substitution Sets of a Form}

In this section , we'll  def\mbox{}ine the termination of the
sequence of the successive substitution sets of a form, which is
directly related to the positive semi-definite property of the form
$F$.

Let $PA=\{A_{[i]}|i=1,2,\cdots,k\}$  be the $L$-substitution
  matrix set.

\begin{definition}\emph{Given the form $F$ on $ \mathbb{T}_n$,  when $[a_1], [a_2],\cdots,[a_m]$ traverse all the memebers of $1,
2,\cdots,k$ respectively, we def\mbox{}ine the set
\begin{displaymath}
\textrm{SLS}^{(m)}(F)=\bigcup\limits_{[\alpha_m]}^{k}\cdots
\bigcup\limits_{[\alpha_{2}]}^{k}\bigcup\limits_{[\alpha_1]}^{k}F(A_{[\alpha_1]}A_{[\alpha_2]}\cdots
A_{[\alpha_m]}X^{\textmd{T}}).
\end{displaymath}
which is called  the $m$-times successive $L$-substitution set of
the form $F$.}
 \end{definition}

\begin{definition}
\emph{  We def\mbox{}ine the sequence of sets
$\{\textrm{SLS}^{(m)}(F)\}_{m=1}^\infty$ as follows}
\begin{displaymath}
\{ \emph{\textrm{SLS}}(F)^{(m)}\}_{m=1}^\infty =
\emph{\textrm{SLS}}(F), \emph{\textrm{SLS}}^{(2)}(F), \cdots .
\end{displaymath}
 \end{definition}

Let $\alpha=(\alpha_1,\alpha_2,\cdots,\alpha_n)\in\mathbb{N}^{n}$,
and let $|\alpha|=\alpha_1+\alpha_2+\cdots \alpha_n$. Then we write
a form $F$ with degree $d$ as $$F=\sum\limits_{|\alpha|=d}c_\alpha
x_1^{\alpha_1}x_2^{\alpha_2}\cdots x_n^{\alpha_n}.$$

 \begin{definition}
\emph{ The form $F$ is called trivially positive if the
coef\mbox{}f\mbox{}icients $c_\alpha$ of every term
$x_1^{\alpha_1}x_2^{\alpha_2}\cdots x_n^{\alpha_n}$ in $F$  are
nonnegative;  If $F(\frac{1}{n},\frac{1}{n},\cdots,\frac{1}{n})<0$,
then $F$ is  called  trivially negative.}
\end{definition}

  \begin{lemma}  \label{zpft}
\emph{Given the form $F$ on $ \mathbb{T}_n$, if the form $F$  is
trivially positive, then  $F\in$ PSD; If the form $F$ is trivially
negative, then  $F~\overline{\in}$ PSD. }
\end{lemma}

\begin{definition}\label{deft}
\emph{Given a form $F$ on
   $\mathbb{T}_n$, if there is a positive integer $k$ such
that every element of the set $\textmd{SLS}^{(k)}(F)$ is trivially
positive, then the sequence of sets
$\{\textrm{SLS}^{(m)}(F)\}_{m=1}^\infty$ is called positively
terminating;  If there is a positive integer $k$ and a form $G$ such
that $G\in \textmd{SLS}^{(k)}(F)$ and $G$ is trivially negative, the
sequence of sets $\{\textrm{SLS}^{(m)}(F)\}_{m=1}^\infty$ is called
negatively terminating;   The sequence of sets
$\{\textrm{SLS}^{(m)}(F)\}_{m=1}^\infty$ is neither positively
terminating nor negatively terminating, then it is called not
terminating.}
 \end{definition}

\section{Nonnegativity decision of forms}

Given a form $F$ on $\mathbb{T}_n$. We know that if the sequence of
sets $\{\textrm{SLS}^{(m)}(F)\}_{m=1}^\infty$ is negatively
terminating, then we can conclude that the form $F$  is positive
semi-definite. Thus there is a natural question, that is, which kind
of forms can be solved by the method? The following theorem answers
the question.

\begin{theorem} \label{zd}
\emph{Let the form $F$ be positive definite on $\mathbb{T}_n$. If
the sequence of the successive $L$-substitution sets is convergent,
then the sequence of sets $\{\textrm{SLS}^{(m)}(F)\}_{m=1}^\infty$
is negatively terminating.}
\end{theorem}
\begin{proof} We only give the proof for the ternary form with
degree $d$, and the multivariate form can be gotten by induction.

Suppose that
$F(x_1,x_2,x_3)=\sum\limits_{i+j+k=d}a_{ijk}x_1^ix_2^jx_3^k$. An
arbitrary $m$-times successive $L$-substitution can be written as
\begin{equation}\label{aaaa}
\left\{ \begin{array}{l} x_1  =  k_{11}u_1+k_{12}u_2+k_{13}u_3,\\
x_2  =  k_{21}u_1+k_{22}u_2+k_{23}u_3,\\
x_3  =  k_{31}u_1+k_{32}u_2+k_{33}u_3.
\end{array}\right.
\end{equation}
Let
\[\begin{array}{lll}
K_1=k_{11}+k_{21}+k_{31}, & K_2=k_{12}+k_{22}+k_{32}, &
K_3=k_{13}+k_{23}+k_{33},\\ t_1=K_1u_1, & t_2=K_2u_2, &
t_3=K_3u_3,\\
 k_1=\frac{k_{11}}{K_1},& k_2=\frac{k_{21}}{K_1}, &
k_3=\frac{k_{31}}{K_1},\\
\alpha_1=\frac{k_{12}}{K_2}-k_1, &  \alpha_2=\frac{k_{22}}{K_2}-k_2,
&  \alpha_3=\frac{k_{32}}{K_2}-k_3,\\
\beta_1=\frac{k_{13}}{K_3}-k_1,& \beta_2=\frac{k_{23}}{K_3}-k_2, &
\beta_3=\frac{k_{33}}{K_3}-k_3.\end{array}\] then (\ref{aaaa})
becomes
\begin{equation}\label{bbbb}
\left\{ \begin{array}{l} x_1  =  k_{1}t_1+(k_{1}+\alpha_1)t_2+(k_{1}+\beta_1)t_3,\\
x_2  =  k_{2}t_1+(k_{2}+\alpha_2)t_2+(k_{2}+\beta_2)t_3,\\
x_3  =  k_{3}t_1+(k_{3}+\alpha_3)t_2+(k_{3}+\beta_3)t_3,
\end{array}\right.
\end{equation}
where $ \sum\limits_{i=1}^3k_i =1, \sum\limits_{i=1}^3\alpha_i = 0$
and $ \sum\limits_{i=1}^3\beta_i=0 $.


Let $t=t_1+t_2+t_3$,  (\ref{bbbb}) can be rewritten as
\begin{equation}
\left\{ \begin{array}{l} x_1  = k_{1}t+\alpha_1t_2+\beta_1t_3,\\
x_2  =  k_{2}t+\alpha_2t_2+\beta_2t_3,\\
x_3  =  k_{3}t+\alpha_3t_2+\beta_3t_3.
\end{array}\right.
\end{equation}
Thus
\begin{equation}
\begin{split}
\Phi(t_1,t_2,t_3) & = F(k_{1}t+\alpha_1t_2+\beta_1t_3,
k_{2}t+\alpha_2t_2+\beta_2t_3,k_{3}t+\alpha_3t_2+\beta_3t_3)\\
 & =
\sum\limits_{i+j+k=d}a_{ijk}(k_{1}t+\alpha_1t_2+\beta_1t_3)^i(k_{2}t+\alpha_2t_2+\beta_2t_3)^j(k_{3}t+\alpha_3t_2+\beta_3t_3)^k \\
&=
\sum\limits_{i+j+k=d}a_{ijk}(k_{1}^it^i+\sum\limits_{p+q+r=i,p\neq
i}\frac{i!}{p!q!r!}k_1^p\alpha_1^q\beta_1^rt^pt_2^qt_3^r)\cdot\\
&(k_{2}^jt^j+\sum\limits_{p+q+r=j,p\neq
j}\frac{j!}{p!q!r!}k_2^p\alpha_2^q\beta_2^rt^pt_2^qt_3^r)(k_{3}^it^i+\sum\limits_{p+q+r=k,p\neq
k}\frac{k!}{p!q!r!}k_3^p\alpha_3^q\beta_3^rt^pt_2^qt_3^r)
\\
&
=\sum\limits_{i+j+k=d}a_{ijk}k_{1}^ik_{2}^jk_{3}^kt^d+\sum\limits_{i+j+k=d}\phi_{ijk}(k_{1},k_{2},k_{3},\alpha_1,\alpha_2,\alpha_3,\beta_1,\beta_2,\beta_3)t_1^it_2^jt_3^k\\
&
=F(k_{1},k_{2},k_{3})t^d+\sum\limits_{i+j+k=d}\phi_{ijk}(k_{1},k_{1},k_{3},\alpha_1,\alpha_2,\alpha_3,\beta_1,\beta_2,\beta_3)t_1^it_2^jt_3^k\\
&
=\sum\limits_{i+j+k=d}\left({\frac{d!}{i!j!k!}}F(k_{1},k_{2},k_{3})+\phi_{ijk}(k_{1},k_{2},k_{3},\alpha_1,\alpha_2,\alpha_3,\beta_1,\beta_2,\beta_3)\right)t_1^it_2^jt_3^k\\
& =\sum\limits_{i+j+k=d}A_{ijk}t_1^it_2^jt_3^k\\
& =\sum\limits_{i+j+k=d}A_{ijk}u_1^iu_2^ju_3^kK_1^iK_2^jK_3^k.
\end{split}
\end{equation}
where
\begin{equation}\label{abds}
\begin{split}
A_{ijk}={\frac{d!}{i!j!k!}}F(k_{1},k_{2},k_{3})+\phi_{ijk}(k_{1},k_{2},k_{3},\alpha_1,\alpha_2,\alpha_3,\beta_1,\beta_2,\beta_3).
\end{split}
\end{equation}

Obviously,
\begin{equation}
\begin{split}
\lim_{(\alpha_1,\alpha_2,\alpha_3,\beta_1,\beta_2,\beta_3)\rightarrow(0,0,0,0,0,0)}\phi_{ijk}(k_{1},k_{2},k_{3},\alpha_1,\alpha_2,\alpha_3,\beta_1,\beta_2,\beta_3)=0.
\end{split}
\end{equation}
And  since  $F(x_1,x_2,x_3)$ is positive def\mbox{}inite on
$\mathbb{T}_n$,  there exists $\varepsilon>0$, such that
\begin{equation}\label{ccc}
F(k_{1},k_{2},k_{3})\geq\varepsilon>0. \end{equation}

On the one hand,   the  vertexs of the subsimplex which corresponds
to the successive $L$-substitution (\ref{bbbb})  are respectively
 $$(k_1,k_2,k_3),
(k_1+\alpha_1,k_2+\alpha_2,k_3+\alpha_3),
(k_1+\beta_1,k_2+\beta_2,k_3+\beta_3).$$ By Theorem (\ref{sl}),
 $\alpha_1,\alpha_2,\alpha_3,\beta_1,\beta_2,\beta_3$ can be suf\mbox{}f\mbox{}iciently small
 when $m$ is  suf\mbox{}f\mbox{}iciently  large.

 On the other hand, for $F$ is continuous on $\mathbb{T}_n$ and by
 (\ref{abds})-(\ref{ccc}), we have $A_{ijk}>
 0$ when $\alpha_1,\alpha_2,\alpha_3,\beta_1,\beta_2,\beta_3$ are suf\mbox{}f\mbox{}iciently
 small.

Putting together the above two aspects,  we have that  there exists
a  suf\mbox{}f\mbox{}iciently  large integer $m$,  such that $F$
becomes trivially positive by (\ref{bbbb}).
 For the successive $L$-substitution (\ref{aaaa}) is arbitrary, the sequence of  sets $\{\textrm{SLS}^{(m)}(F)\}_{m=1}^\infty$ is
positively terminating. \end{proof}

 Let $L$ be the Yang-Yao's substitution set in Theorem \ref{zd}, then the theorem  is the main result
 in
 \cite{Yong1}.  However,   the proof of the theorem in this paper is dif\mbox{}ferent from  the one given in
\cite{Yong1}.

According to the proof of Theorem  \ref{zd}, we obtain the following
conclusion.

\begin{corollary}\label{zd111}
\emph{Let the form $F$ be positive definite on $\mathbb{T}_n$. If
the sequence of the successive $L$-substitution sets is convergent,
then by an arbitrary $m$-times successive $L$-substitution, when $m$
is suf\mbox{}f\mbox{}iciently larger,  $F$ can become a nonlacunary
form whose coef\mbox{}f\mbox{}icients are all
 positive.} \label{gzd}
\end{corollary}

 In Theorem \ref{zd}, if
the sequence of the successive $L$-substitution sets isn't
convergent, the conclusion of the  theorem  is not always true.

\begin{theorem} \label{zd1}
\emph{Let the form $F$ be positive definite on $\mathbb{T}_n$. If
the sequence of the successive $L$-substitution sets isn't
convergent,  the sequence of sets
$\{\textrm{SLS}^{(m)}(F)\}_{m=1}^\infty$ isn't always negatively
terminating.}
\end{theorem}

\begin{proof}   Let $L$ be the substitution set,  which correspands to the   subdivision of the simplex
$\mathbb{T}_3$  given by Fig.4(a). And we have concluded that the
sequence of the successive $L$-substitution sets isn't convergent.

Given the form
 \begin{equation}
\begin{split}
F(x_1,x_2,x_3)=(x_1-x_2+x_3)^2+x_2^2, \quad (x_1,x_2,x_3)\in
\mathbb{T}_3.
\end{split}
\end{equation}
Apparently, $F$ is positive definite on $\mathbb{T}_n$.  Consider
the  $m$-times successive $L$-substitution
$$X^{\textmd{Tr}}=A_{[\alpha]}^mT^{\textmd{Tr}},$$
where
\begin{equation} A_{[\alpha]}= \left[
\begin{array}{lll}
1 & 0 & \frac{1}{3} \\
0 & 1 & \frac{1}{3}\\
0 & 0 & \frac{1}{3}
\end{array} \right].
\end{equation}
And without difficulty, we have
\begin{equation}
\begin{split}
A_{[\alpha]}^m& = \left[\begin{array}{rrr}
\frac{3 }{2}& -\frac{1 }{2} &1 \\
\frac{1 }{2}& -\frac{1 }{2} &0\\
0 & 1 & 0
\end{array}\right]
\left[\begin{array}{lll}
1& 0 &0 \\
0 & \frac{1}{3^m} & 0\\
0 & 0 & 1
\end{array}\right] \left[\begin{array}{rrr}
\frac{3 }{2}& -\frac{1 }{2} &1 \\
\frac{1 }{2}& -\frac{1 }{2} &0\\
0 & 1 & 0
\end{array}\right]^{-1}\\
& =\left[\begin{array}{ccc}
1& 0 &\frac{1}{2}(1-\frac{1}{3^m}) \\
0& 1 &\frac{1}{2}(1-\frac{1}{3^m}) \\
0 & 0 & \frac{1}{3^m}
\end{array}\right],
\end{split}
\end{equation}
Then
\begin{equation}
\begin{split}
 F(A_{[\alpha]}^m\cdot(x_1,x_2,x_3)^{\textmd{T}}) = & x_1^2-2x_1x_2
 +\frac{2}{3^m}x_1x_3+2x_2^2+(1-\frac{1}{3^{m-1}})x_2x_3+\\
 & \frac{1}{4}(\frac{5}{3^{2m}}-\frac{2}{3^{m}}+1)x_3^2.
 \end{split}
\end{equation}
When $m\longrightarrow\infty$,  the coef\mbox{}f\mbox{}icient of the
term $x_1x_2$ is -2.  Therefore,  the sequence of sets
$\{\textrm{SLS}^{(m)}(F)\}_{m=1}^\infty$ isn't terminating.
\end{proof}

\section{Decision of indef\mbox{}inite forms}
Many problems,  such  as the inequality disproving, are always
transformed into decision of indef\mbox{}inite forms. Given a form
$F$ on $\mathbb{T}_n$. Suppose that there exists $X_0\in
\mathbb{T}_n$ such that $F(X_0)>0$. It is well-known to us that if
the sequence of sets $\{\textrm{SLS}^{(m)}(F)\}_{m=1}^\infty$ is
negatively terminating, then the form $F$  is indef\mbox{}inite.
Then it follows a question naturally: for a indef\mbox{}inite form
$F$ on $\mathbb{T}_n$, is the
$\{\textrm{SLS}^{(m)}(F)\}_{m=1}^\infty$ negatively terminating? The
question is answered by the following theorem.

\begin{theorem} \label{ce}
\emph{Let the form $F$ be indef\mbox{}inite on $\mathbb{T}_n$. If
the sequence of the successive $L$-substitution sets is convergent,
then the sequence of sets $\{\textrm{SLS}^{(m)}(F)\}_{m=1}^\infty$
is negatively terminating.}
\end{theorem}
\begin{proof} Let $PA$ be the  substitution
  matrix set which corresponds to the substitution set  $L$.  Since the form $F$ is indef\mbox{}inite on
   $\mathbb{T}_n$,  there exists $X_0\in
\mathbb{T}_n$ such that $F(X_0)<0$ .  And $F$ is continuous on
$\mathbb{T}_n$,  so there exists a neighborhood
$\textrm{U}(X_0)\subset \mathbb{T}_n$ of $X_0$ (If $X_0$ is on the
boundary of $\mathbb{T}_n$, then we take $\textrm{U}(X_0)\bigcap
\mathbb{T}_n$) such that  $F(X)<0$ for all $X\in \textrm{U}(X_0)$.
For the sequence of the successive $L$-substitution sets is
convergent, then there exists a subsimplex $\sigma$  of the $k$-th
 subdivision of $\mathbb{T}_n$, which corresponds to the
$k$-times successive $L$-substitution
 $$X^{\textmd{Tr}}=B_{[i_1]}B_{[i_2]}\cdots B_{[i_k]}T^{\textmd{Tr}},~~ B_{[i_1]}, B_{[i_2]}, \cdots,B_{[i_k]}\in PW_n,$$
 satisfying $\sigma
\subset\textrm{U}(X_0)$, where $k$ is a  suf\mbox{}f\mbox{}iciently
larger integer. Thus, $-F(X)$ is positive definite on
 $\sigma$. By Theorem  \ref{zd},
the sequence of  sets $\{\textrm{SLS}^{(m)}(-F)\}_{m=1}^\infty$ is
positively terminating, so there exists a
 $l$-times successive
$L$-substitution
 $$X^{\textmd{Tr}}=B_{[j_1]}B_{[j_2]}\cdots
B_{[j_l]}T^{\textmd{Tr}},  ~~ B_{[j_1]}, B_{[j_2]},
\cdots,B_{[j_l]}\in PW_n,$$
 satisfying that $F(B_{[i_1]}B_{[i_2]}\cdots B_{[i_k]}B_{[j_l]}B_{[j_2]}\cdots B_{[j_l]}T^{\textmd{Tr}})$
 is trivially negative. Therefore, the sequence of sets $\{\textrm{SLS}^{(m)}(F)\}_{m=1}^\infty$ is
negatively terminating.\end{proof}

By the proving process of Theorem (\ref{ce}),  we obtain the
following algorithm,  which is used to decide the indef\mbox{}inite
form with a counter-example.

\textbf{Algorithm} \textbf{(SLS)}\\
Input: the form  $F\in \mathbb{Q}[x_1,x_2,\cdots, x_n]$, where  $F$ is positive def\mbox{}inite or  indef\mbox{}inite on $\mathbb{R}_{+}^n$.  \\
Output:  ``$F\in $PSD'', or ``$F(\tilde{X}_0)<0$''.\\
step1: Let  $\mathbb{F}=\{F\}$.\\
step2:  Compute $\bigcup\limits_{F\in\mathbb{F}}\textrm{SLS}(F)$,

\qquad Let
$$\mathbb{F}=\bigcup\limits_{F\in\mathbb{F}}\textrm{SLS}(F)-\{\textmd{trivially
positive forms in}
\bigcup\limits_{F\in\mathbb{F}}\textrm{SLS}(F)\}\triangleq\{F_{[1]},F_{[2]},\cdots,F_{[k]}\},$$

\qquad where
$$F_{[i]}= F(B_{[i]}X^{\textmd{Tr}}),\quad B_{[i]} \in PW_n.$$
 step3: Let $I=[[1],[2],\cdots,[k]]$.

\quad step31:  If $\mathbb{F}$ is null, then output ``$F\in $PSD'',
and terminate.

\quad step32:  If there is a trivially negative form
$F_{[i]}\in\mathbb{F}$, then output $``F(\tilde{X}_0)<0\mbox{''}$,

~~~~~~~~~~~~ and
  terminate, where
 $$\tilde{X}_0 =
 B_{I[i]}(\frac{1}{n},\frac{1}{n},...,\frac{1}{n})^{\textmd{Tr}}, ~B_{I[i]}=B_{[I[i][1]]}B_{[I[i][2]]}\cdots
B_{[I[i][m]]},$$  \quad \quad \quad \quad\qquad

~~~~~~~~~~~~  $I[i]$ is the $i$-th component of $I$, $I[i][j]$ is
the $j$-th component
 of $I[i]$, and $m$ is the

  ~~~~~~~~~~~~  the total number of the components of $I[i]$.

\quad step33:  Else, Compute
$\bigcup\limits_{F\in\mathbb{F}}\textrm{SLS}(F)$.  Let

\begin{displaymath}
\begin{split}
\mathbb{F} &
=\bigcup\limits_{F\in\mathbb{F}}\textrm{SLS}(F)-\{\textmd{trivially
positive forms in}
\bigcup\limits_{F\in\mathbb{F}}\textrm{SLS}(F)\}\\
& =
\{F_{[\textrm{op}(I[1]),1]},\cdots,F_{[\textrm{op}(I[1]),l_1]},F_{[\textrm{op}(I[2]),1]},\cdots,F_{[\textrm{op}(I[2]),l_2]},\\
&
\cdots,F_{[\textrm{op}(I[k]),1]},\cdots,F_{[\textrm{op}(I[k]),l_k]}\},
\end{split}
\end{displaymath}

~~~~~~~~~~~~  where
$$F_{[\textrm{op}(I[i]),j]}=F(B_{[\textrm{op}(L[i]),j]}X^{\textmd{Tr}}),$$

~~~~~~~~~~~~  and $\textrm{op}(I[i])$ extracts operands from
$\textrm{op}(I[i])$. And let
\begin{displaymath}
\begin{split}
L& =[[\textrm{op}(I[1]),1],\cdots,[\textrm{op}(I[1]),l_1],[\textrm{op}(I[2]),1],\cdots,[\textrm{op}(I[2]),l_2],\\
& \cdots,[\textrm{op}(I[k]),1],\cdots,[\textrm{op}(I[k]),l_k]],
\end{split}
\end{displaymath}

~~~~~~~~~~~~then go to step3.

By Algorithm SLS, we design  a Maple program called SLS, see
Appendix. To the program SLS,  there are some positive semi-definite
forms making the program do not terminate, that is, we cann't decide
these positive semi-definite forms by the method.

\section{Examples}

In this section, we demonstrate the program  SLS with some examples
on a computer with  Intel(R) Core(TM)2 Duo CPU (E7200 @ 2.53GHz) and
3.25G RAM.

Let  $L_1,L_2,L_3$ be the substitution sets   which correspond to
the subdivisions of the simplex $\mathbb{T}_3$  given by Fig.1,
Fig.2(b) and Fig.2(c), respectively. And let  $PA_i$ be the
$L_i$-substitution matrix set for $i=1,2,3$.

 \textbf{Example 1.} Show that the following form is positive semi-definite on
 $\mathbb{R}_{+}^3$,
 \begin{equation}
F(x,y,z)=x(x-y)^5-y(-z-y)^5-z(x-z)^5 .
\end{equation}

 Utilize the program SLS and execute order SLS$(F,PA_i,[x,y,z])$ for $i=1, 2, 3$.   And the running
 results are shown in Table 1. So $F$ positive semi-definite on
 $\mathbb{R}_{+}^3$.

 \begin{table}[H]   
 \begin{center}
\setlength{\arrayrulewidth}{0.1mm}
\renewcommand{\arraystretch}{1.2}
\begin{tabular}{|c|c|c|c|}
 \hline
 The substitution set & Running SLS
times & Output   & CPU time (s)  \\
 \hline
$L_1$ & 3 & $F\in $PSD  & 0.016  \\
 \hline
$L_2$ & 3 & $F\in $PSD  & 0.015 \\
 \hline
$L_3$ & 2 &$F\in $PSD  & 0.032  \\
 \hline

\multicolumn{4}{c} {Table ~1.} \\
\end{tabular}
\end{center}
\end{table}

\textbf{Example 2.}  Show that the following form is
indef\mbox{}inite on $ \mathbb{R}_{+}^3$,
\begin{equation}\label{eq2}
F(x,y,z)=x^4y^2-2x^4yz+x^4z^2+3x^3y^2z-2x^3yz^2-2x^2y^4-2x^2y^3z+x^2y^2z^2+y^6,
\end{equation}

Execute order SLS$(F,PA_i,[x,y,z])$ for $i=1, 2, 3$, and the running
 results are shown in Table 2. Obviously, $F(0,1,0)=1>0$, so the  form $F$ is indef\mbox{}inite on $
\mathbb{R}_{+}^3$.

\begin{table}[H]   
\begin{center}
\setlength{\arrayrulewidth}{0.1mm}
\renewcommand{\arraystretch}{1.2}
\begin{tabular}{|c|c|c|c|}
 \hline
 The  substitution set & Running SLS
times & Output   & CPU time (s)  \\
 \hline
$L_1$ & 3 & $F(\frac{37}{81}, \frac{91}{324}, \frac{85}{324})<0$  & 0.094 \\
 \hline
$L_2$ & 3 & $F(\frac{11}{24}, \frac{1}{3}, \frac{5}{24})<0$ & 0.062 \\
 \hline
$L_3$ & 2 & $F(\frac{13}{27}, \frac{7}{27}, \frac{7}{27})<0$  & 0.094  \\
 \hline
\multicolumn{4}{c} {Table ~2.} \\
\end{tabular}
\end{center}
\end{table}

\textbf{Example 3.}   Prove or disprove that, for $x\geq 0, y\geq 0,
z\geq 0$, and $x+y+z\neq 0$,
\begin{equation}\label{eq1}
 \frac{2}{3}(\frac{x^2}{y+z}+\frac{y^2}{z+x}+\frac{z^2}{x+y})-(\frac{x^6+y^6+z^6}{3})^{\frac{1}{6}}\geq
0,
\end{equation}

Take of\mbox{}f denominators of the left polynomial, and  denote the
new polynomial by $F$. Execute order SLS$(F,PA_i,[x,y,z])$ for $i=1,
2, 3$, the running
 results are shown in Table 3. So the  inequality doesn't hold.

\begin{table}[H]   
\begin{center}
\renewcommand{\arraystretch}{1.2}
\setlength{\arrayrulewidth}{0.1mm}
\begin{tabular}{|c|c|c|c|}
 \hline
  The  substitution set & Running SLS
times & Output   & CPU time (s)  \\
 \hline
$L_1$ & 5 & $F(\frac{2159}{5832}, \frac{3685}{11664}, \frac{3661}{11664})<0$  & 582.750  \\
 \hline
$L_2$ & 5 & $F(\frac{7}{24}, \frac{37}{96}, \frac{31}{96})<0$
 & 21.469 \\
 \hline
$L_3$ & 3 & $F(\frac{31}{81}, \frac{25}{81}, \frac{25}{81})<0$  & 234.484  \\
 \hline

\multicolumn{4}{c} {Table~ 3.} \\
\end{tabular}
\end{center}
\end{table}
The examples above indicate  the effectiveness of Algorithm SLS.
 Thus, we obtain various effective
 substitutions  for deciding positive semi-definite forms
and indefinite forms which are beyond Yang's substitutions
characterized by ``difference''.

\subsection*{Appendix.  Maple Program SLS}
\begin{maplettyout}
 SLS:=proc(poly,A,var)
 local f,i,j,k,m,n,p,s,t,F,G,H,M,newvar,st,Var:
 uses combinat, LinearAlgebra:
 t:=time(): F:=[[poly,[0]]]:n:=nops(var):r£º=100£º
 Var:=convert(var,Vector):
 for i to nops(A) do
     for j to n do
        newvar[i,j]:=op(j,convert(A[i].Var,list)):
     od:
 od:
 for s to r do
     m:=nops(F): f:=[]:
     for k to m do
         G[k]:=[]:
         for i to nops(A) do
             st:={seq(Var[j]=newvar[i,j],j=1..n)}:
             G[k]:=[op(G[k]),[expand(subs(st,F[k][1])),[op(F[k][2]),i]]]:
         od:
     od:
     F:=[seq(op(G[u]),u=1..nops(F))]:
     for i to nops(F) do
         if max([coeffs(F[i][1])]) < 0 then
            print(F[i][2]):
            M :=IdentityMatrix(n):
            for j from 2 to nops(F[i][2]) do
                M:=M.A[F[i][2][j]]:
            od:
            print(convert(M.Vector[column](n,1/n),list)):
            print(time()-t,`second`):
            return ("The form is indefinite"):
        elif min([coeffs(F[i][1])]) >=0 then
             f:=[op(f),i]:
        fi:
     od:
     if nops(f)>0 then
        F:=subs({seq(F[f[p]]=NULL,p=1..nops(f))},F):
     fi:
     if nops(F)=0 then
        print(s):print(time()-t,`second`):
        return("The form is positive semi-definite"):
     fi:
 od:
 end proc:
\end{maplettyout}
 \clearpage
\end{document}